\documentclass{article}
\usepackage{amsmath}

\oddsidemargin  - 5 mm
\evensidemargin - 5 mm
\newtheorem{theorem}{Theorem}

\newtheorem{proposition}[theorem]{Proposition}

\input{tcilatex}

\begin{document}

\title{Elliptic Curves and Algebraic Geometry Approach in Gravity Theory II.
Parametrization of a Multivariable Cubic Algebraic Equation}
\author{Bogdan G. Dimitrov \thanks{%
Electronic mail: bogdan@theor.jinr.ru} \\
Bogoliubov Laboratory for Theoretical Physics\\
Joint Institute for Nuclear Research \\
6 Joliot -Curie str. \\
Dubna 141980, Russia}
\maketitle

\begin{abstract}
\ \ In a previous paper, the general approach for treatment of algebraic
equations of different order in gravity theory was exposed, based on the
important distinction between covariant and contravariant metric tensor
components.

In the present second part of the paper it has been shown that a
multivariable cubic algebraic equation can also be parametrized by means of
complicated, irrational and non-elliptic functions, depending on the
elliptic Weierstrass function and its derivative. As a model example, the
proposed before cubic algebraic equation for reparametrization invariance of
the gravitational Lagrangian has been investigated.

This is quite different from the standard algebraic geometry approach, where
only the parametrization of two-dimensional cubic algebraic equations have
been considered. Also, the possible applications in modern cosmological
theories has been commented.
\end{abstract}

\section{\protect\bigskip INTRODUCTION}

Recently there has been an increasing interest in finding solutions of the
Einstein's equations in terms of elliptic and theta functions. In [1] a
solution of the Einstein's equations with a perfect fluid energy-momentum
tensor was found for the special case of the inhomogeneous Szafron -
Szekeres cosmological model [2, 3,4,5 ] with a metric 
\begin{equation}
ds^{2}=dt^{2}-e^{2\alpha (t,r,y,z)}dr^{2}-e^{2\beta (t,r,y,z)}(dy^{2}+dz^{2})%
\text{ \ \ \ }  \tag{1.1}
\end{equation}%
Solutions expressed through ultraelliptic functions for the case of an
relativistic gravitational field of a rigidly rotating disk of dust have
been found in [6, 7, 8]. Further, in [1] important cosmological
characteristics for observational cosmology such as the \textit{Hubble's
constant} $H(t)=\frac{\overset{.}{R}(t)}{R(t)}$ and the \textit{deceleration
parameter} $q=-\frac{\overset{..}{R}(t)R(t)}{\overset{.}{R}^{2}(t)}$ have
been expressed in terms of the \textit{Jacobi's theta function} and of the 
\textit{Weierstrass elliptic function} respectively. Thus solutions in terms
of elliptic functions may turn out to be an important ingredient in
understanding the s. c. \textquotedblright accelerating cosmological
models\textquotedblright .

Since in [1] a solution has been found only for one component of the
Einstein's equations and also for one additional, limiting case - the FLRW
(Friedman - Lemaitre - Robertson - Walker) metric, it is important to know
whether there are other cosmological metrics, for which solutions in terms
of elliptic functions exist.

In [9, 10], instead of seaching new solutions for special cases, a more
general approach has been proposed. Concretely, it has been shown that the
Einstein's vacuum equations and the gravitational Lagrangian can be
represented in the form of multivariable cubic algebraic equations. In [9]
this was performed for a specific choice of the contravariant tensor
components in the form $\widetilde{g}^{ij}=dX^{i}dX^{j}$, where $%
X^{i}=X^{i}(x_{1},x_{2},..,x_{n})$ are some generalized coordinates,
depending on the initial coordinates $x_{1},x_{2},..,x$. In [10], this was
performed for a general contravariant metric tensor \ $\widetilde{g}^{ij}$.
In both papers, it has been assumed that the contravariant metric components
differ from the inverse one (i.e. $g_{ij}\widetilde{g}^{jk}=l_{i}^{k}\neq
\delta _{i}^{k}$) and consequently, a \textit{distinction should be made
between covariant and contravariant metric tensor components.} This is the
essence of the \textit{affine geometry approach}, the essence of which have
been clarified in the previous paper ([10], Part I). 

Most importantly, the derived cubic equations in all cases clearly suggest
that in respect to a chosen variable, they can be brought to the
parametrizable form of the cubic algebraic equation [11, 12, 13] 
\begin{equation}
y^{2}=4x^{3}-g_{2}x-g_{3}\text{ \ \ \ \ \ \ .\ }  \tag{1.2}
\end{equation}%
Then, in accordance with the standard algebraic geometry prescriptions, one
can set up 
\begin{equation}
x=\rho (z)\text{ \ ; \ \ }y=\rho ^{^{\prime }}(z)\text{ \ \ \ ,}  \tag{1.3}
\end{equation}%
$g_{2}=60\sum\limits_{\omega \subset \Gamma }\frac{1}{\omega ^{4}}$ and $%
g_{3}=140\sum\limits_{\omega \subset \Gamma }\frac{1}{\omega ^{6}}$ are
complex numbers, called the \textit{Eisenstein series} and $\rho (z)$
denotes the \textit{Weierstrass elliptic function} $\rho (z)=\frac{1}{z^{2}}%
+\sum\limits_{\omega }\left[ \frac{1}{(z-\omega )^{2}}-\frac{1}{\omega ^{2}}%
\right] $ , where the summation is over the poles $\omega $ in the complex
plane.

\textbf{\ }Previously \ [9], it has been shown how a selected variable $%
dX^{i}$ of the cubic algebraic equation 
\begin{equation}
dX^{i}dX^{l}\left( p\Gamma _{il}^{r}g_{kr}dX^{k}-\Gamma
_{ik}^{r}g_{lr}d^{2}X^{k}-\Gamma _{l(i}^{r}g_{k)r}d^{2}X^{k}\right)
-dX^{i}dX^{l}R_{il}=0\text{ \ \ \ \ }  \tag{1.4}
\end{equation}%
of \textit{reparametrization invariance of the gravitational Lagrangian} \
and the ratio $\frac{a}{c}$ of \ two coefficient functions in the performed
linear - fractional transformation can be parametrized with the Weierstrass
function and its derivative.

The main purpose of this second part of the paper will be to demonstrate how
this parametrization can be extended to all the variables $dX^{i}$\ in the
cubic algebraic equation and thus a non-plane, multivariable cubic algebraic
equation will also be possible to be parametrized, but in terms of more
complicated functions. This is necessary to be performed in order to find
how the contravariant metric components $\widetilde{g}^{ij}=dX^{i}dX^{j} $
are parametrized - the importance of this will become evident in the next
(third) part of this paper. \textbf{\ }

It will be demonstrated in section 2 how \textit{an embedded system of cubic
algebraic equations} can be obtained, and each equation from this sequence
will be for the algebraic subvariety of the solutions - i.e. the first
equation will be for the algebraic variety of $n$ variables, the second one
- for $(n-1)$ variables,...., the last one will be only for the $dX^{1}$
variable\textbf{. }In section 3 it has been proved that yet the situation is
different from the standard case - the obtained functions for the
differentials $dX^{1}$, $dX^{2}$, $dX^{3}$ are complicated irrational
functions of the Weierstrass function $\rho (z)$, its derivative $\rho
^{^{\prime }}(z)$, the (covariant)\ metric tensor and affine connection
components, which depend on the global coordinates $X^{1}$, $X^{2}$, $X^{3}$%
. Therefore, it might seem that due to the dependence on the global
coordinates one cannot assert that the functions $dX^{1}$, $dX^{2}$, $dX^{3}$
are \textquotedblright uniformization functions\textquotedblright , since
they should depend only on the complex coordinate $z$.

In the next (third) part of this paper it shall be proved that yet the
uniformization functions for the cubic equation can be found since the
generalized coordinates $X^{1}$, $X^{2}$, $X^{3}$ can be expressed as
solutions of a first - order system of nonlinear differential equations and
thus the dependence of the differentials on $X^{1}$, $X^{2}$, $X^{3}$ is
eliminated.

Throughout the whole paper it has been assumed that the second differentials
of the generalized coordinates are zero $d^{2}X^{i}=0$. This may seem to be
a serious restriction, but it was necessary to be imposed in order to be
able to construct the algorithm for finding the solutions of the cubic
equation. The important point is another - since the main tool of the
proposed approach is the \textit{linear - fractional transformation}, will
this approach be applicable also for the more complicated cases of 1. the
cubic algebraic equation of reparametrization invariance, but with $%
d^{2}X^{i}\neq 0$ and 2. the system of cubic and quartic Einstein's
algebraic equations. In the subsequent parts of this paper, it will be shown
that the approach continues to work, although with some modifications and
technical difficulties

\section{\ EMBEDDED \ SEQUENCE \ OF \ ALGEBRAIC \ EQUATIONS \ AND \ FINDING
\ THE \ SOLUTIONS \ OF \ \ THE \ CUBIC \ ALGEBRAIC \ EQUATION}

The purpose of the present subsection will be to describe the method for
finding the solution (i. e . the algebraic variety of the differentials $%
dX^{i}$) of the cubic algebraic equation (1.4) (in the limit $d^{2}X^{k}=0$%
). The applied method has been proposed first in [9] but here it will be
developed further and applied with respect to a sequence of algebraic
equations with algebraic varieties, which are embedded into the initial one.
This means that if at first the algorithm is applied with respect to the
three-dimensional cubic algebraic equation (1.4) and a solution for $dX^{3}$
(depending on the Weierstrass function and its derivative is found), then
the same algorithm will be applied with respect to the two-dimensional cubic
algebraic equation with variables $dX^{1}$\ and $dX^{2}$, and finally to the
one-dimensional cubic algebraic equation of the variable $dX^{1}$\ only.

The basic and very simple idea about parametrization of a cubic algebraic
equation with the Weierstrass function [11-13] can be presented as follows:
Let us define the lattice $\Lambda =\{m\omega _{1}+n\omega _{2}\mid m,n\in Z;
$ $\omega _{1},\omega _{2}\in C,Im\frac{\omega _{1}}{\omega _{2}}>0\}$ and
the mapping $f:$ $C/\Lambda \rightarrow CP^{2}$, which maps the factorized
(along the points of the lattice $\Lambda $) part of the points on the
complex plane into the two dimensional complex projective space $CP^{2}$.
This means that each point $z$ on the complex plane is mapped onto the point 
$(x,y)=(\rho (z),\rho ^{^{\prime }}(z))$, where $x$ and $y$ belong to the
affine curve (1.2). In other words, the functions $x=\rho (z)$ and $y=\rho
^{^{\prime }}(z)$ are \textit{uniformization functions} for the cubic curve
\ and it can be proved [13] that the only cubic algebraic curve (but with
number coefficients!) which is parametrized by the uniformization functions $%
x=\rho (z)$ and $y=\rho ^{^{\prime }}(z)$ is the above mentioned affine
curve.

In the case of the cubic equation (1.4), the aim will be again to bring the
equation to the form (1.2) and afterwards to make equal each of the
coefficient functions to the (numerical) coefficients in (1.2).

In order to provide a more clear description of the developed method, let us
divide it into several steps.

\textbf{Step 1}. The initial cubic algebraic equation (1.4) is presented as
a cubic equation with respect to the variable $dX^{3}$ only 
\begin{equation}
A_{3}(dX^{3})^{3}+B_{3}(dX^{3})^{2}+C_{3}(dX^{3})+G^{(2)}(dX^{2},dX^{1},g_{ij},\Gamma _{ij}^{k},R_{ik})\equiv 0%
\text{ \ \ \ \ ,}  \tag{2.1}
\end{equation}%
where naturally the coefficient functions $A_{3}$, $B_{3}$ , $C_{3}$ and $%
G^{(2)}$ depend on the variables $dX^{1}$ and $dX^{2}$ of the algebraic
subvariety and on the metric tensor $g_{ij}$, the Christoffel connection $%
\Gamma _{ij}^{k}$ and the Ricci tensor $R_{ij}$: 
\begin{equation}
A_{3}\equiv 2p\Gamma _{33}^{r}g_{3r}\text{ \ \ ; \ \ \ \ \ \ \ \ \ \ \ }%
B_{3}\equiv 6p\Gamma _{\alpha 3}^{r}g_{3r}dX^{\alpha }-R_{33}\text{\ \ \ \
,\ }  \tag{2.2}
\end{equation}%
\begin{equation}
C_{3}\equiv -2R_{\alpha 3}dX^{\alpha }+2p(\Gamma _{\alpha \beta
}^{r}g_{3r}+2\Gamma _{3\beta }^{r}g_{\alpha r})dX^{\alpha }dX^{\beta }\text{
\ \ \ .}  \tag{2.3}
\end{equation}%
The Greek indices $\alpha ,\beta $ take values $\alpha ,\beta =1,2$ while
the indice $r$ takes values $r=1,2,3$.

\textbf{Step 2}. A linear-fractional transformation 
\begin{equation}
dX^{3}=\frac{a_{3}(z)\widetilde{dX}^{3}+b_{3}(z)}{c_{3}(z)\widetilde{dX}%
^{3}+d_{3}(z)}  \tag{2.4 }
\end{equation}%
is performed with the purpose of setting up to zero the coefficient
functions in front of the highest (third) degree of $\ \widetilde{dX}^{3}$.
This will be achieved if $G^{(2)}(dX^{2},dX^{1},g_{ij},\Gamma
_{ij}^{k},R_{ik})=-\frac{a_{3}Q}{c_{3}^{3}}$, where 
\begin{equation}
Q\equiv A_{3}a_{3}^{2}+C_{3}c_{3}^{2}+B_{3}a_{3}c_{3}+2c_{3}d_{3}C_{3}\text{
\ \ \ \ \ \ , }  \tag{2.5 }
\end{equation}%
which gives a cubic algebraic equation with respect to the two-dimensional
algebraic variety of the variables $dX^{1}$ and $dX^{2}$: 
\begin{equation}
p\Gamma _{\gamma (\alpha }^{r}g_{\beta )r}dX^{\gamma }dX^{\alpha }dX^{\beta
}+K_{\alpha \beta }^{(1)}dX^{\alpha }dX^{\beta }+K_{\alpha }^{(2)}dX^{\alpha
}+2p\left( \frac{a_{3}}{c_{3}}\right) ^{3}\Gamma _{33}^{r}g_{3r}=0\text{ \ \
\ }  \tag{2.6 }
\end{equation}%
and $K_{\alpha \beta }^{(1)}$ and $K_{\alpha }^{(2)}$ are the corresponding
quantities [9] 
\begin{equation}
K_{\alpha \beta }^{(1)}\equiv -R_{\alpha \beta }+2p\frac{a_{3}}{c_{3}}(1+2%
\frac{d_{3}}{c_{3}})(2\Gamma _{\alpha \beta }^{r}g_{3r}+\Gamma _{3\alpha
}^{r}g_{\beta r})\text{ \ \ \ }  \tag{2.7 }
\end{equation}%
and 
\begin{equation}
K_{\alpha }^{(2)}\equiv 2\frac{a_{3}}{c_{3}}\left[ 3p\frac{a_{3}}{c_{3}}%
\Gamma _{\alpha 3}^{r}g_{3r}-(1+2\frac{d_{3}}{c_{3}})R_{\alpha 3}\right] 
\text{ \ \ .}  \tag{2.8}
\end{equation}%
Note that since the linear fractional transformation (with another
coefficient functions) will again be applied with respect to another cubic
equations, everywhere in (2.4 - 2.7) the coefficient functions $a_{3}(z)$, $%
b_{3}(z)$, $c_{3}(z)$ and $d_{3}(z)$ bear the indice \textquotedblright $3$%
\textquotedblright ,\ to distinguish them from the indices in the other
linear-fractional tranformations. In terms of the new variable $n_{3}=%
\widetilde{dX}^{3}$ the original cubic equation (1.4)\ acquires the form [9] 
\begin{equation}
\widetilde{n}^{2}=\overline{P}_{1}(\widetilde{n})\text{ }m^{3}+\overline{P}%
_{2}(\widetilde{n})\text{ }m^{2}+\overline{P}_{3}(\widetilde{n})\text{ }m+%
\overline{P}_{4}(\widetilde{n})\text{ ,}  \tag{2.9 }
\end{equation}%
where $\overline{P}_{1}(\widetilde{n})$ $,\overline{P}_{2}(\widetilde{n}),$ $%
\overline{P}_{3}(\widetilde{n})$ and $\overline{P}_{4}(\widetilde{n})$ are
complicated functions of the ratios $\frac{c_{3}}{d_{3}}$, $\frac{b_{3}}{%
d_{3}}$ and $A_{3},B_{3},C_{3}$ (but not of the ratio $\frac{a_{3}}{d_{3}}$,
which is very important). The variable $m$ denotes the ratio $\frac{a_{3}}{%
c_{3}}$ and the variable $\widetilde{n}$ is related to the variable $n$
through the expresssion 
\begin{equation}
\widetilde{n}=\sqrt{k_{3}}\sqrt{C_{3}}\left[ n+L_{1}^{(3)}\frac{B_{3}}{C_{3}}%
+L_{2}^{(3)}\right] \text{ \ \ \ \ \ ,}  \tag{2.10}
\end{equation}%
where 
\begin{equation}
k_{3}\equiv \frac{b_{3}}{d_{3}}\frac{c_{3}}{d_{3}}(\frac{c_{3}}{d_{3}}+2)%
\text{ \ \ \ \ \ ,}  \tag{2.11}
\end{equation}%
\begin{equation}
L_{1}^{(3)}\equiv \frac{1}{2}\frac{\frac{b_{3}}{d_{3}}}{\frac{c_{3}}{d_{3}}+2%
}\text{ \ \ ; \ \ \ }L_{2}^{(3)}\equiv \frac{1}{\frac{c_{3}}{d_{3}}+2}\text{
\ \ .}  \tag{2.12}
\end{equation}

The subscript \textquotedblright $3$\textquotedblright\ in $L_{1}^{(3)}$ and 
$L_{2}^{(3)}$ means that the corresponding ratios in the R. H. S. also have
the same subscript. Setting up the coefficient functions $\overline{P}_{1}(%
\widetilde{n})$ $,\overline{P}_{2}(\widetilde{n}),$ $\overline{P}_{3}(%
\widetilde{n})$ equal to the number coefficients $4,0,-g_{2},-g_{3}$
respectively, one can now parametrize the resulting equation 
\begin{equation}
\widetilde{n}^{2}=4m^{3}-g_{2}m-g_{3}\text{ }  \tag{2.13 }
\end{equation}%
according to the standard prescription 
\begin{equation}
\widetilde{n}=\rho ^{^{\prime }}(z)=\frac{d\rho }{dz}\text{ \ \ \ \ \ \ \ \
\ \ \ ;\ \ \ \ \ \ \ \ \ \ \ \ \ \ }\frac{a_{3}}{c_{3}}\equiv \text{\ }%
m=\rho (z)\text{\ \ \ \ .}  \tag{2.14 }
\end{equation}%
Taking this into account, representing the linear-fractional transformation
(2.4) as (dividing by $\ c_{3}$) 
\begin{equation}
dX^{3}=\frac{\frac{a_{3}}{c_{3}}\widetilde{dX}^{3}+\frac{b_{3}}{c_{3}}}{%
\widetilde{dX}^{3}+\frac{d_{3}}{c_{3}}}  \tag{2.15 }
\end{equation}%
and combining expressions (2.10) for $\widetilde{n}$ and (2.15), one can
obtain the final formulae for $dX^{3}$ as a solution of the cubic algebraic
equation 
\begin{equation}
dX^{3}=\frac{\frac{b_{3}}{c_{3}}+\frac{\rho (z)\rho ^{^{\prime }}(z)}{\sqrt{%
k_{3}}\sqrt{C_{3}}}-L_{1}^{(3)}\frac{B_{3}}{C_{3}}\rho (z)-L_{2}^{(3)}\rho
(z)}{\frac{d_{3}}{c_{3}}+\frac{\rho ^{^{\prime }}(z)}{\sqrt{k_{3}}\sqrt{C_{3}%
}}-L_{1}^{(3)}\frac{B_{3}}{C_{3}}-L_{2}^{(3)}}\text{ \ \ \ \ \ .}  \tag{2.
16}
\end{equation}%
In order to be more precise, it should be mentioned that the identification
of the functions $\overline{P}_{1}(\widetilde{n})$ $,\overline{P}_{2}(%
\widetilde{n}),$ $\overline{P}_{3}(\widetilde{n})$ with the number
coefficients gives some additional equations [9], which in principle have to
be taken into account in the solution for $dX^{3}$. This has been
investigated to a certain extent in [9], and will be continued to be
investigated. Here in this paper the main objective will be to show the
dependence of the solutions on the Weierstrass function and its derivative.
Since only the ratios $\frac{b}{d}$ and $\frac{c}{d}$ enter these additional
relations, and not $\frac{a}{c}$ (which is related to the Weierstrass
function), they do not affect the solution with respect to $\rho (z)$ and $%
\rho ^{^{\prime }}(z)$.

Since $B_{3}$ and $C_{3}$ depend on $dX^{1}$ and $dX^{2}$, the solution
(2.16)\ for $dX^{3}$ shall be called \textit{the embedding solution} for $%
dX^{1}$ and $dX^{2}$.

\textbf{Step 3.} Let us now consider the two-dimensional cubic equation
(2.6). Following the same approach and finding the \textquotedblright
reduced\textquotedblright\ cubic algebraic equation for $dX^{1}$ only, it
shall be proved that the solution for $dX^{2}$ is the embedding solution for 
$dX^{1}$.

For the purpose, let us again write down eq. (2.6)\ in the form (2.1),
singling out the variable $dX^{2}$: 
\begin{equation}
A_{2}(dX^{2})^{3}+B_{2}(dX^{2})^{2}+C_{2}(dX^{2})+G^{(1)}(dX^{1},g_{ij},%
\Gamma _{ij}^{k},R_{ik})\equiv 0\text{ \ \ \ \ ,}  \tag{2.17}
\end{equation}%
where the coefficient functions $A_{2},B_{2},C_{2}$ and $G^{(1)}$ are the
following: 
\begin{equation}
A_{2}\equiv 2p\Gamma _{22}^{r}g_{2r}\text{ \ \ ; \ \ \ \ \ \ \ \ \ \ \ }%
B_{2}\equiv K_{22}^{(1)}+2p[2\Gamma _{12}^{r}g_{2r}+\Gamma
_{22}^{r}g_{1r}]dX^{1}\text{\ \ \ \ ,\ }  \tag{2.18 }
\end{equation}%
\begin{equation}
C_{2}\equiv 2p[\Gamma _{11}^{r}g_{2r}+2\Gamma
_{12}^{r}g_{1r})(dX^{1})^{2}+(K_{12}^{(1)}+K_{21}^{(1)})dX^{1}+K_{2}^{(2)}%
\text{ \ \ \ ,}  \tag{2.19 }
\end{equation}%
\begin{equation}
G^{1}\equiv 2p\Gamma
_{11}^{r}g_{1r}(dX^{1})^{3}+K_{11}^{(1)}(dX^{1})^{2}+K_{1}^{(2)}dX^{1}+2p%
\rho ^{3}(z)\Gamma _{33}^{r}g_{3r}\text{ \ \ .}  \tag{2.20 }
\end{equation}%
Note that the starting equation (2.6) has the same structure of the first
terms, if one makes the formal substitution $-R_{\alpha \beta }\rightarrow
K_{\alpha \beta }^{(1)}$ in the second terms, but eq. (2.6)\ has two more
additional terms $K_{1}^{(2)}dX^{1}+2p\rho ^{3}(z)\Gamma _{33}^{r}g_{3r}.$
Therefore, one might guess how the coefficient functions will look like just
by taking into account the above substitution and the contributions from the
additional terms. Revealing the general structure of the coefficient
functions might be particularly useful in higher dimensions, when one would
have a \textquotedblright chain\textquotedblright\ of cubic algebraic
equations. Concretely for the three-dimensional case, investigated here, $%
C_{2}$ in (2.19) can be obtained from $C_{3}$ in (2.3), observing that there
will be an additional contribution from the term $K_{\alpha
}^{(2)}dX^{\alpha }$ for $\alpha =2$. Also, in writing down the coefficient
functions in (2.1)\ it has been accounted that as a result of the previous
parametrization $\frac{a_{3}}{c_{3}}=\rho (z)$ .

Since eq. (2.17)\ is of the same kind as eq. (2.1), for which we already
wrote down the solution, the expression for $dX^{2}$ will be of the same
kind as in formulae (2.16), but with the corresponding functions $%
A_{2},B_{2},C_{2}$ instead of $A_{3},B_{3},C_{3}$. Taking into account (2.18
- 2.19), the solution for $dX^{2}$ can be written as follows: 
\begin{equation}
dX^{2}=\frac{\frac{1}{\sqrt{k_{2}}}\rho (z)\rho ^{^{\prime }}(z)\sqrt{C_{2}}%
+h_{1}(dX^{1})^{2}+h_{2}(dX^{1})+h_{3}}{\frac{1}{\sqrt{k_{2}}}\rho
^{^{\prime }}(z)\sqrt{C_{2}}+l_{1}(dX^{1})^{2}+l_{2}(dX^{1})+l_{3}}\text{ \
\ \ , }  \tag{2.21}
\end{equation}%
where $h_{1},h_{2},h_{3},l_{1},l_{2},l_{3}$ are expressions, depending on $%
\frac{b_{2}}{d_{2}},\frac{d_{2}}{c_{2}},\Gamma _{\alpha \beta }^{r}$ ($%
r=1,2,3$ ; $\alpha ,\beta =1,2$), $g_{\alpha \beta }$, $K_{12}^{(1)}$, $%
K_{21}^{(1)}$ and on the Weierstrass function. They will be presented in
Appendix $A$.

The representation of the solution for $dX^{2}$ in the form (2.21)\ shows
that it is \textit{an "embedding" solution} of $dX^{1}$ in the sense that it
depends on this function.Correspondigly, the solution (2.16) for $dX^{3}$ is
an \textit{"embedding"} one for the variables $dX^{1}$ and $dX^{2}$.

\textbf{Step 4.} It remains now to investigate the one-dimensional cubic
algebraic equation 
\begin{equation}
A_{1}(dX^{1})^{3}+B_{1}(dX^{1})^{2}+C_{1}(dX^{1})+G^{(0)}(g_{ij},\Gamma
_{ij}^{k},R_{ik})\equiv 0\text{ \ \ \ \ ,}  \tag{2.22 }
\end{equation}%
obtained from the two-dimensional cubic algebraic equation (2.17)\ after
applying the linear-fractional transformation 
\begin{equation}
dX^{2}=\frac{a_{2}(z)\widetilde{dX}^{2}+b_{2}(z)}{c_{2}(z)\widetilde{dX}%
^{2}+d_{2}(z)}  \tag{2.23 }
\end{equation}%
and setting up to zero the coefficient function before the highest (third)\
degree of $(dX^{2})^{3}$. Taking into account that as a result of the
previous parametrization $\frac{a_{2}}{c_{2}}=\rho (z)$ , the coefficient
functions $A_{1},B_{1},C_{1}$and $D_{1}$ are given in a form, not depending
on $dX^{2}$ and $dX^{3}$: 
\begin{equation}
A_{1}\equiv 2p\Gamma _{11}^{r}g_{1r}\text{ \ \ \ ,}  \tag{2.24 }
\end{equation}%
\begin{equation}
B_{1}\equiv F_{3}\rho (z)+K_{11}^{(1)}=2p(1+2\frac{d_{2}}{c_{2}})[2\Gamma
_{12}^{r}g_{1r}+\Gamma _{11}^{r}g_{2r}]\rho (z)+K_{11}^{(1)}\text{ \ \ ,} 
\tag{2.25 }
\end{equation}%
\begin{equation*}
C_{1}\equiv F_{1}\rho ^{2}(z)+F_{2}\rho (z)+K_{1}^{(2)}=2p[2\Gamma
_{12}^{r}g_{2r}+\Gamma _{22}^{r}g_{1r}]\rho ^{2}(z)+
\end{equation*}%
\begin{equation}
+(1+2\frac{d_{2}}{c_{2}})(K_{12}^{(1)}+K_{21}^{(1)})\rho (z)+K_{1}^{(2)}%
\text{ \ \ \ \ \ ,}  \tag{2.26 }
\end{equation}%
\begin{equation}
G^{0}\equiv 2p[\Gamma _{22}^{r}g_{2r}+\Gamma _{33}^{r}g_{3r}]\rho
^{3}(z)+K_{22}^{(1)}\rho ^{2}(z)\text{ \ \ \ \ .}  \tag{2.27 }
\end{equation}%
The solution for $dX^{1}$ can again be written in the form (2.16), but with $%
\frac{b_{1}}{c_{1}}$, $\frac{d_{1}}{c_{1}}$, $L_{1}^{(1)}$, $L_{2}^{(1)}$, $%
k_{1}$and $B_{1},C_{1}$ instead of these expressions with the indice
\textquotedblright $3$\textquotedblright .

Taking into account formulaes (2.24 - 2.27)\ for $A_{1},B_{1}$ and $C_{1}$,
the final expression for $dX^{1}$ can be written as 
\begin{equation}
dX^{1}=\frac{\frac{1}{\sqrt{k_{1}}}\rho (z)\rho ^{^{\prime }}(z)\sqrt{%
F_{1}\rho ^{2}+F_{2}\rho (z)+K_{1}^{(2)}}+f_{1}\rho ^{3}+f_{2}\rho
^{2}+f_{3}\rho +f_{4}}{\frac{1}{\sqrt{k_{1}}}\rho ^{^{\prime }}(z)\sqrt{%
F_{1}\rho ^{2}(z)+F_{2}\rho (z)+K_{1}^{(2)}}+\widetilde{g}_{1}\rho ^{2}(z)+%
\widetilde{g}_{2}\rho (z)+\widetilde{g}_{3}}\text{ \ \ \ \ ,}  \tag{2.28}
\end{equation}%
where $F_{1},F_{2},f_{1},f_{2},f_{3},f_{4},\widetilde{g}_{1},\widetilde{g}%
_{2}$ and $\widetilde{g}_{3}$ are functions (also to be given in Appendix $A$%
), depending on $g_{\alpha \beta }$, $\Gamma _{\alpha \beta }^{r}$ ($\alpha
,\beta =1,2$) and on the ratios $\frac{b_{1}}{c_{1}}$, $\frac{b_{1}}{d_{1}}$%
, $\frac{b_{2}}{d_{2}}$, $\frac{d_{1}}{c_{1}}$, $\frac{d_{2}}{c_{2}}$.

\section{\protect\bigskip A \ PROOF \ THAT \ THE \ SOLUTIONS \ $%
dX^{1},dX^{2} $ AND $dX^{3}$ ARE \ NOT \ ELLIPTIC \ FUNCTIONS}

\begin{proposition}
The expressions (2.21) for $dX^{2}$ and (2. 28) for $dX^{1}$do not represent
elliptic functions.
\end{proposition}

Proof: The proof is straightforward and will be based on assuming the
contrary. Let us first assume that $dX^{1}$ is an elliptic function. Then
from standard theory of elliptic functions [11] it follows that $dX^{1}$%
(being an elliptic function by assumption) can be represented as 
\begin{equation}
dX^{1}=K_{1}(\rho )+\rho ^{^{\prime }}(z)K_{2}(\rho )\text{ \ \ \ \ ,} 
\tag{3.1}
\end{equation}%
where $K_{1}(\rho )$ and $K_{2}(\rho )$ depend on the Weierstrass function
only. For convenience one may denote the expressions outside the square root
in the nominator and denominator as 
\begin{equation}
Z_{1}(\rho )\equiv f_{1}\rho ^{3}(z)+f_{2}\rho ^{2}(z)+f_{3}\rho (z)+f_{4}%
\text{ \ \ \ ,}  \tag{3.2}
\end{equation}%
\begin{equation}
Z_{2}(\rho )\equiv g_{1}\rho ^{2}(z)+g_{2}\rho (z)+g_{3}\text{ \ \ \ \ \ \ .}
\tag{3.3}
\end{equation}%
Then, setting up equal the expressions (3.1)\ and (2.28) for $dX^{1}$, one
can express the function $K_{2}(\rho )$ as 
\begin{equation}
K_{2}(\rho )=\frac{(1-K_{1}(\rho ))\rho \rho ^{^{\prime }}\sqrt{F_{1}\rho
^{2}+F_{2}\rho +K_{1}^{(2)}}+\sqrt{k_{1}}Z_{1}(\rho )-Z_{2}(\rho )K_{1}(\rho
)}{\rho ^{^{\prime }}\left[ \rho \sqrt{F_{1}\rho ^{2}+F_{2}\rho +K_{1}^{(2)}}%
+\sqrt{k_{1}}Z_{2}(\rho )\right] }\text{ \ \ .}  \tag{3.4}
\end{equation}%
The R. H. S. of the above expression depends on the derivative $\rho
^{^{\prime }}$, while the L.H. S. depends on $\rho $ only. Therefore the
obtained contradiction is due to the initial assumption that $dX^{1}$ is an
elliptic function, having the representation (3.1). In quite a similar way,
it can be proved that $dX^{2}$ and $dX^{3}$ are not elliptic functions. In
fact, this is obvious since they are embedding functions of $dX^{1}$, which
is not an elliptic function.

\section{\protect\bigskip DISCUSSION}

The most important result in this paper is related to the possibility to
find the parametrization functions for a multicomponent cubic algebraic
surface, again by consequent application of the linear-fractional
transformation. The parametrization functions represent complicated
irrational expressions of the Weierstrass function and its derivative,
unlike in the standard two - dimensional case, where they are the
Weierstrass function itself and its first derivative. The advantage of
applying the linear- fractional transformations (2.4) and (2.23) is that by
adjusting their coefficient functions (so that the highest - third degree in
the transformation equation will vanish), the following sequence of plane
cubic algebraic equations is fulfilled (the analogue of eq.(65) in [9]): 
\begin{equation}
P_{1}^{(3)}(n_{(3)})m_{(3)}^{3}+P_{2}^{(3)}(n_{(3)})m_{(3)}^{2}+P_{3}^{(3)}(n_{(3)})m_{(3)}+P_{(4)}^{(3)}=0%
\text{ \ \ \ \ \ ,}  \tag{4.1}
\end{equation}%
\begin{equation}
P_{1}^{(2)}(n_{(2)})m_{(2)}^{3}+P_{2}^{(2)}(n_{(2)})m_{(2)}^{2}+P_{3}^{(2)}(n_{(2)})m_{(2)}+P_{(4)}^{(2)}=0%
\text{ \ \ \ \ \ ,}  \tag{4.2}
\end{equation}%
\begin{equation}
P_{1}^{(1)}(n_{(1)})m_{(1)}^{3}+P_{2}^{(1)}(n_{(1)})m_{(1)}^{2}+P_{3}^{(1)}(n_{(1)})m_{(1)}+P_{(4)}^{(1)}=0%
\text{ \ \ \ \ \ ,}  \tag{4.3}
\end{equation}%
where $m_{(3)}$, $m_{(2)}$, $m_{(1)}$ denote the ratios $\frac{a_{3}}{c_{3}}$%
, $\frac{a_{2}}{c_{2}}$, $\frac{a_{1}}{c_{1}}$ in the corresponding linear -
fractional transformations and $n_{(3)}$, $n_{(2)}$, $n_{(1)}$ are the
\textquotedblright new\textquotedblright\ variables $\widetilde{dX}^{3}$, $%
\widetilde{dX}^{2}$, $\widetilde{dX}^{1}$. The sequence of plane cubic
algebraic equations (4.1 - 4.3) should be understood as follows: the first
one (4.1) holds if the second one (4.2) is fulfilled; the second one (4.2)
holds if the third one (4.3) is fulfilled. Of course, in the case of $n$
variables (i. e. $n$ component cubic algebraic equation) the generalization
is straightforward. Further, since each one of the above plane cubic curves
can be transformed to the algebraic equation ($i=1,2,3$) 
\begin{equation}
\widetilde{n}_{(i)}^{2}=\overline{P}_{1}^{(i)}(\widetilde{n}%
_{(i)})m_{(i)}^{3}+\overline{P}_{2}^{(i)}(\widetilde{n}_{(i)})m_{(i)}^{2}+%
\overline{P}_{3}^{(i)}(\widetilde{n}_{(i)})m_{(i)}+\overline{P}_{4}^{(i)}(%
\widetilde{n}_{(i)})  \tag{4.4}
\end{equation}%
and subsequently to its parametrizable form, one obtains the solutions of
the initial multicomponent cubic algebraic equation.

\section{APPENDIX \ A: \ SOME \ COEFFICIENT \ FUNCTIONS \ IN \ THE \ FINAL \
SOLUTIONS \ FOR \ $dX^{1}$, $dX^{2}$, $dX^{3}$ \ IN \ SECTION \ 2}

\bigskip The functions $h_{1},h_{2},h_{3}$ (depending on the Weierstrass
function $\rho (z)$) and the functions $l_{1},l_{2},l_{2}$ (not depending on 
$\rho (z)$) in the expression (2.23) for the solution $dX^{2}$ of the cubic
algebraic equation are 
\begin{equation}
h_{1}\equiv 2p\left( \frac{b_{2}}{c_{2}}-2\frac{b_{2}}{d_{2}}L_{1}^{(2)}\rho
(z)\right) \left( 2\Gamma _{12}^{r}g_{1r}+\Gamma _{11}^{r}g_{2r}\right) 
\text{ \ \ \ \ ,}  \tag{A1}
\end{equation}%
\begin{equation*}
h_{2}\equiv \left( \frac{b_{2}}{c_{2}}-2\frac{b_{2}}{d_{2}}L_{1}^{(2)}\rho
(z)\right) \left( K_{12}^{(1)}+K_{21}^{(1)}\right) -
\end{equation*}%
\begin{equation}
-2pL_{1}^{(2)}\rho (z)\left( 2\Gamma _{12}^{r}g_{2r}+\Gamma
_{22}^{r}g_{1r}\right) \text{ \ \ \ \ \ ,}  \tag{A2}
\end{equation}%
\begin{equation}
h_{3}\equiv \left( \frac{b_{2}}{c_{2}}-2\frac{b_{2}}{d_{2}}L_{1}^{(2)}\rho
(z)\right) K_{2}^{(2)}-L_{1}^{(2)}\rho (z)K_{22}^{(1)}\text{ \ \ \ ,} 
\tag{A3}
\end{equation}%
\begin{equation}
l_{1}\equiv 2p\left( \frac{d_{2}}{c_{2}}-2\frac{b_{2}}{d_{2}}%
L_{1}^{(2)}\right) \left( 2\Gamma _{12}^{r}g_{1r}+\Gamma
_{11}^{r}g_{2r}\right) \text{ \ \ \ \ ,}  \tag{A4}
\end{equation}%
\begin{equation*}
l_{2}\equiv \left( \frac{d_{2}}{c_{2}}-2\frac{b_{2}}{d_{2}}%
L_{1}^{(2)}\right) \left( K_{12}^{(1)}+K_{21}^{(1)}\right) -
\end{equation*}%
\begin{equation}
-2pL_{1}^{(2)}\left( 2\Gamma _{12}^{r}g_{2r}+\Gamma _{22}^{r}g_{1r}\right) 
\text{ \ \ \ \ \ ,}  \tag{A5}
\end{equation}%
\begin{equation}
l_{3}\equiv \left( \frac{d_{2}}{c_{2}}-2\frac{b_{2}}{d_{2}}%
L_{1}^{(2)}\right) K_{2}^{(2)}-L_{1}^{(2)}K_{22}^{(1)}\text{ \ \ \ .} 
\tag{A6}
\end{equation}%
The functions $F_{1},F_{2},F_{3},f_{1},f_{2},f_{3},f_{4},\widetilde{g}_{1},%
\widetilde{g}_{2},\widetilde{g}_{3}$ in the solution for $dX^{1}$ are the
following 
\begin{equation}
F_{1}\equiv 2p\left( 2\Gamma _{12}^{r}g_{2r}+\Gamma _{22}^{r}g_{1r}\right) 
\text{ \ \ ; \ }F_{2}\equiv \left( 1+2\frac{d_{2}}{c_{2}}\right) \left(
K_{12}^{(1)}+K_{21}^{(1)}\right) \text{ \ \ ,}  \tag{A7}
\end{equation}%
\begin{equation}
F_{3}\equiv 2p\left( 1+2\frac{d_{2}}{c_{2}}\right) \left( 2\Gamma
_{12}^{r}g_{1r}+\Gamma _{11}^{r}g_{2r}\right) \text{ \ \ \ \ ,}  \tag{A8}
\end{equation}%
\begin{equation}
f_{1}\equiv -2\frac{b_{1}}{d_{1}}L_{1}^{(1)}F_{1}\text{ \ \ ; \ \ }%
f_{3}\equiv \frac{b_{1}}{c_{1}}F_{2}-L_{1}^{(1)}\text{ \ ,}  \tag{A9}
\end{equation}%
\begin{equation}
f_{2}\equiv \frac{b_{1}}{c_{1}}F_{1}-L_{1}^{(1)}F_{3}-2\frac{b_{1}}{d_{1}}%
L_{1}^{(1)}F_{2}\text{ \ \ ,}  \tag{A10}
\end{equation}%
\begin{equation}
\widetilde{g}_{1}\equiv \left( \frac{d_{1}}{c_{1}}-2\frac{b_{1}}{d_{1}}%
\right) F_{1}\text{ \ \ ,}  \tag{A11}
\end{equation}%
\begin{equation}
\widetilde{g}_{2}\equiv \left( \frac{d_{1}}{c_{1}}-2\frac{b_{1}}{d_{1}}%
\right) F_{2}-L_{1}^{(1)}F_{3}\text{ \ \ ,}  \tag{A12}
\end{equation}%
\begin{equation}
\widetilde{g}_{3}\equiv \left( \frac{d_{1}}{c_{1}}-2\frac{b_{1}}{d_{1}}%
\right) K_{1}^{(2)}-L_{1}^{(1)}K_{11}^{(1)}\text{ \ \ \ .}  \tag{A13}
\end{equation}

\section{Acknowledgments}

The author is grateful to Dr. L. K. Alexandrov, St. Mishev (BLTP, JINR,
Dubna) and especially to Prof. V. V. Nesterenko, Dr. O. Santillan (BLTP,
JINR, Dubna) and to Prof. Sawa Manoff (INRNE, Bulgarian Academy of Sciences,
Sofia) for valuable comments, discussions and critical remarks. \ 

This paper is written in memory of \ Prof. S. S. Manoff (1943 - 27.05.2005)
- a specialist in classical gravitational theory. \ 

The author is grateful also to Dr. A. Zorin (LNP, JINR) and to J. Yanev
(BLTP, JINR) for various helpful advises and to Dr.V. Gvaramadze (Sternberg
Astronomical Institute, MSU, Moscow) and his family for their moral support
and encouragement.

\end{document}